\documentclass[11pt,letterpaper]{article}

\usepackage{eurosym}
\usepackage{graphicx}
\usepackage{times}

\begin{document}

\vspace*{-2.5cm}

\begin{center}
{\LARGE\bf Altruism may arise from individual selection}\\[8mm]

{\large Angel S\'anchez\footnote{\mbox{Corresponding author.
Phone: +34-916\,249\,411. 
Fax: +34-916\,249\,129. 
{\tt anxo@math.uc3m.es}}}
and Jos\'e A. Cuesta}\\[4mm]
{\large\em
Grupo Interdisciplinar de Sistemas Complejos (GISC)\\[0mm]
Departamento de Matem\'aticas\\[0mm]
Universidad Carlos III de Madrid\\[0mm]
28911 Legan\'es, Madrid, Spain}

\thispagestyle{empty}

\vspace{1cm}
{\large\bf Abstract}
\end{center}

The fact that humans cooperate with non-kin in large groups, or with people 
they will never meet again, is a long-standing evolutionary puzzle. 
Altruism, the capacity to perform costly acts that confer benefits on
others, is at the core of cooperative behavior. 
Behavioral experiments show that humans have a predisposition to cooperate with
others and to punish non-cooperators at personal cost (so-called strong reciprocity)
which, 
according to standard evolutionary game arguments,
can not arise from selection acting on 
individuals. This has led to the suggestion of group and cultural
selection as the only mechanisms that can explain the evolutionary
origin of human altruism. 
We introduce an agent-based model inspired on the Ultimatum
Game, that allows us to go beyond the limitations of standard evolutionary
game theory and show that 
individual selection can indeed give rise to strong reciprocity.
Our results are consistent with the existence of neural correlates
of fairness and in good agreement with observations on humans 
and monkeys. 

\vspace*{1cm}

\noindent
{\bf Keywords:} Strong reciprocity, Individual selection,
Evolutionary theories,\\
\phantom{\bf Keywords:} Behavioral evolution, Evolutionary game theory

\newpage

\section{Introduction}

Ever since Darwin first faced this problem 
(Darwin, 1871; Gould, 2002),
the arising of human cooperation has been a subject of intense 
debate within the framework of evolutionary theories. 
Cooperation has been linked to altruism, which can be defined as
the capacity to perform costly acts that confer benefits on
others 
(Fehr and Fischbacher, 2003).
Previous theoretical approaches to altruism have shown that 
in many instances altruistic behavior is not truly so, in so far 
as they 
yield benefits for the altruist in the future.
This is the case when the recipients of the altruistic act are 
relatives, well understood within kin selection theory 
(Hamilton, 1964).
Altruism in the absence of kin relationships has also been 
explained in terms of repeated interaction leading to cooperation
(Axelrod and Hamilton, 1981;
Trivers, 1971),
indirect benefit through 
reputation gains
(Leimar and Hammerstein, 2001;
Milinski et al., 2002;
Nowak and Sigmund, 1998)
or costly signalling theories 
(Gintis et al., 2001).
However, recent behavioral experiments show that humans can 
perform altruistic acts when interactions are anonymous and 
one-shot, i.e., in conditions which exclude all the explanations
proposed so far 
(Fehr et al., 2002; Fehr and G\"achter, 2002;
Fehr and Rockenbach, 2003; Henrich et al., 2001).
Indeed, it has been observed that individuals are ready to punish 
non-cooperators (altruistic punishment) as well as to reward 
cooperative behavior (altruistic rewarding) even when doing so 
will not produce any benefit for the punisher or rewarder. 
This set of behaviors has been termed 
strong reciprocity 
(Fehr et al., 2002; Gintis, 2000)
and, as such, it 
has been proposed as a schema for understanding
altruism in humans 
(Fehr and Fischbacher, 2003; Gintis et al., 2003).

Substantial evidence in favor of the existence of strong 
reciprocity comes from experiments using the so-called 
Ultimatum Game 
(G\"uth et al., 1982),
and from agent-based models 
(Bowles et al., 2003b; Bowles and Gintis, 2004; Boyd et al., 2003)
[see
(Fehr and Fischbacher, 2003; Gintis et al., 2003)
for summaries].
In the Ultimatum Game,
under conditions of anonymity, two players are shown a sum 
of money, say 100 \EUR{}. One of the players, the ``proposer'',
is instructed to offer any amount, from 1 \EUR{} to 100 \EUR{},
to the other, the ``responder''. The proposer can make only one
offer, which the responder can accept or reject. If the offer is
accepted, the money is shared accordingly; if rejected, both 
players receive nothing. Since the game is played only once 
(no repeated interactions) and anonymously (no reputation gain),
a self-interested responder will accept any amount of money 
offered. Therefore, self-interested proposers will offer the 
minimum possible amount, 1 \EUR{}, which will be accepted. 
%
%
To be sure, this is a backward-induction way of reasoning, which leads
to the conclusion that the subgame-perfect Nash equilibrium is the 
relevant one. However, the Ultimatum game has many Nash equilibria, 
which can play a role in the results we report below (see, e.g., 
Samuelson, 1997, or Gintis, 2000, for complete game-theoretical 
discussions on this issue). We will come
back to this question in Sec.\ 6. 
%
%
Notwithstanding, 
in actual Ultimatum Game experiments with human subjects, 
average offers do not even approximate the self-interested prediction.
Generally speaking, proposers offer respondents very substantial
amounts (50 \% being a typical modal offer) and respondents
frequently reject offers below 30 \%. Most of the 
experiments have been carried out with university students in
western countries, showing a large degree of individual variability
but a striking uniformity between groups in average behavior.
%
%
A recent experiment (G\"uth et al., 2003) used newspaper readers in order to have a 
population with broader characteristics and background,
%
%
finding 
qualitatively similar results.
%
%
Interestingly, 
%
%
a large study in 15 small-scale societies 
(Henrich et al., 2001)
found that, in all
cases, respondents or proposers behave in a reciprocal manner. 
Furthermore, the behavioral variability across groups was much
larger than previously observed: while mean offers in the case 
of university students are in the range 43\%-48\%, in the 
cross-cultural study they ranged from 26\% to 58\%. 

The fact that indirect reciprocity is excluded by the anonymity 
condition and that interactions
are one-shot allows one to interpret rejections in terms of 
strong reciprocity 
(Fehr et al., 2002; Gintis, 2000).
This amounts to considering that
these behaviors are truly altruistic, i.e., that 
they are costly for the individual performing them in so far as 
they do not result in direct or indirect benefit. As a consequence, 
we immediately face an evolutionary puzzle: the negative effects of
altruistic acts must decrease the altruist's fitness as compared to
the that of the recipients of the benefit, ultimately leading to 
the extinction of altruists. Indeed, standard evolutionary game 
theory arguments applied to the Ultimatum Game lead to the expectation
that in a mixed population, punishers (individuals 
who reject low offers) have less chance to survive 
%
%
than
%
%
rational 
players (indivuals who accept any offer) and eventually disappear
(Page and Nowak 2000, 2002).
Although 
much attention has been devoted to this issue by 
researchers in differents aspects of evolutionary theory, 
the problem is yet far from understood
(Bowles et al., 2003a; Hammerstein, 2003; Vogel, 2004)
To date, the only way out to 
this dilemma seems, following the original suggestion of 
Darwin 
(Darwin, 1871),
to invoke
group and cultural
selection to compensate for the negative effects
that reciprocity is assumed to have on individuals
(Bowles et al., 2003b, Boyd et al., 2003; Hammerstein, 2003).

\section{One parameter model}

In order to assess the possible evolutionary origins of these 
behaviors, we introduce and analyze here a drastically simplified model. 
Imagine a population of $N$ players of the Ultimatum Game 
with a fixed sum of money $M$ per game. 
Random pairs of players are chosen, of which one is the proposer
and another one is the respondent. In its simplest version,
we will assume that 
players are capable of other-regarding behavior (empathy); consequently,
in order to optimize their gain,
proposers offer the minimum amount of money 
that they would accept. Every agent has her own, fixed 
acceptance threshold, $1\leq t_i\leq M$ ($t_i$ are always integer
numbers for simplicity). Agents have only one strategy:
respondents reject any offer 
smaller than their own acceptance threshold, and 
accept offers otherwise.
Although we believe that this is the way 
in which `empathic' agents will behave, in order not to hinder 
other strategies {\em a priori}, we have also considered
the possibility that agents have two
independent acceptance and offer thresholds. As we will see 
below, this
does not change our main results and conclusions.
Money
shared as a consequence of accepted offers accumulates to the 
capital of each of the involved players. As our main aim is to 
study selection acting on modified descendants, hereafter we interpret this 
capital as `fitness' 
(here used in a loose, Darwinian sense, not in the more 
restrictive one of reproductive rate).
After $s$ games,
the agent with the overall minimum fitness is 
removed (randomly picked if there are several)
and a new agent is introduced by duplicating that
with the maximum fitness, i.e., with the same threshold and the
same fitness (again randomly picked if there are 
several). Mutation is introduced in the duplication process by 
allowing changes of $\pm 1$ in the acceptance threshold of the
newly generated player with probability 1/3 each. Agents 
have no memory (i.e., interactions are one-shot) and no information
about other agents (i.e., no reputation gains are possible).

%
%
Two remarks about our model are in order before proceeding any 
further. First, we need to clarify the motivation for our choice 
of simple, memoryless agents. 
It is likely that in early human societies some degree of repeated 
interaction and reputation effects was present, factors that we have
excluded from our model. In this respect, we stress that what
we are actually trying to show is that the behavior observed in the
experiments quoted above (Fehr and Fischbacher, 2003), 
can arise by individual selection in the absence of precisely those
two ingredients, repeated interactions and reputation: In other 
words, the existence of repeated interactions and reputation is not
a necessary condition for the selection of altruistic-like behaviors
at the individual level. In that case, actual circumstances of human
evolution would reinforce the tendency to the appearance of altruism.
The fact that similar results are found in Ultimatum game
experiments in a wide range of small scale societies
(Henrich et al., 2001)
suggests that our conclusions will have to be kept in mind when 
dealing with early human behavior, as the relevance of these two 
influences is largely different in the studied societies.
Second, we want to stress that our mutation rate, which we choose 
somewhat large to enhance the fluctuation effects (see related
comments in Sec.\ \ref{sec:two} below), should not be understood from
the genetic viewpoint, but rather from the phenotypical viewpoint. 
Indeed, the inheritance of an acceptance threshold like the one we
are proposing may perfectly be also affected by cultural transmission, 
and it is therefore subject to a large individual variability. 
Observations reported in the literature (Fehr and Fischbacher, 2003;
Gintis et al., 2003) support this great variability. 
On the other hand, it has to be borne in mind that even if
the mutation rate may seem large, mutations are small, with relative
changes of the order of 1/100 in the acceptance threshold. We believe
that such changes from parent to child are actually very likely, and
hence our choice for the mutation rate. 
%
%

\section{Results}

\begin{figure}
\begin{center}
\includegraphics[height=14cm,angle=270]{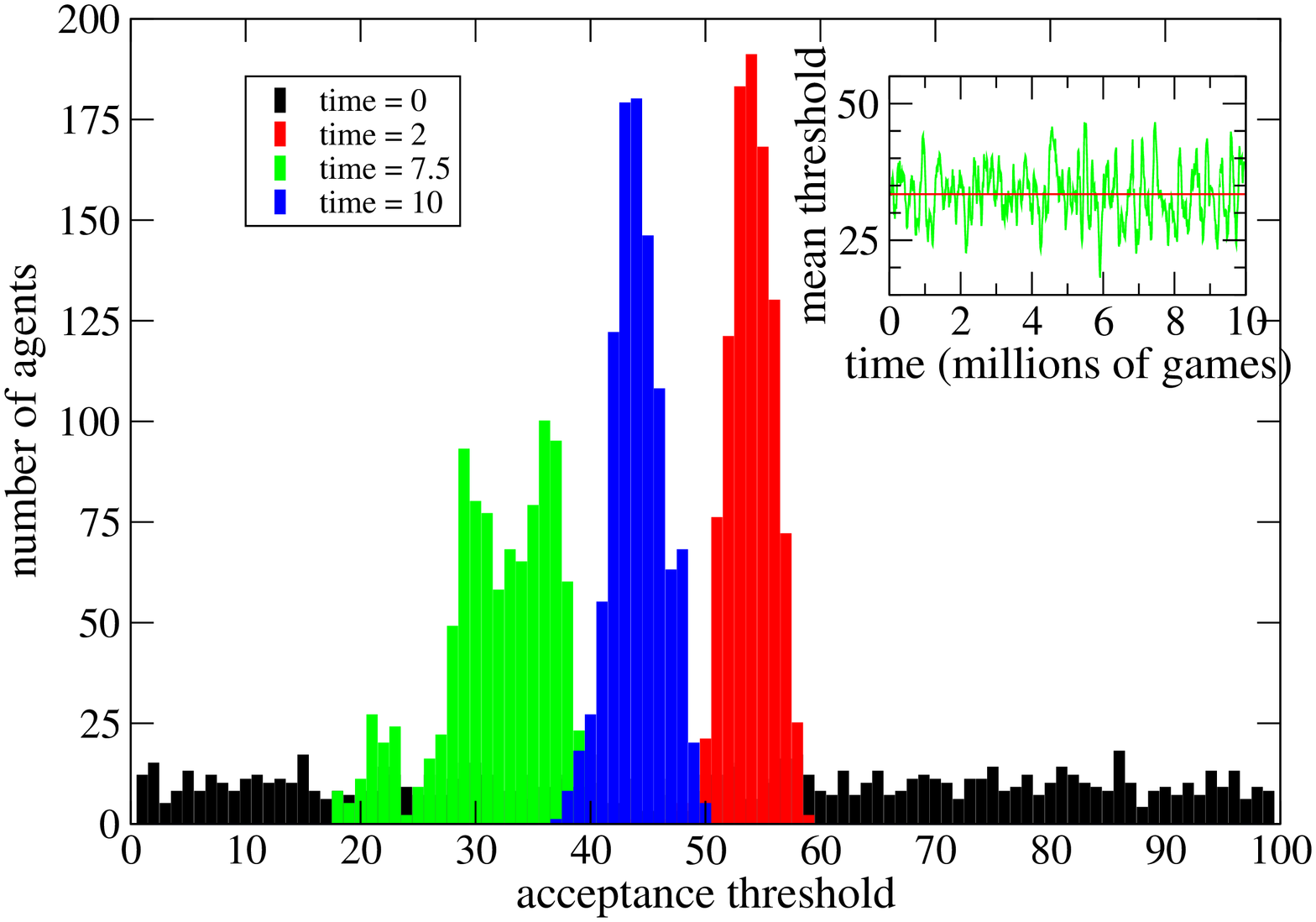}
\caption{\label{fig:smallpop}
}
\end{center}
\end{figure}

Figure \ref{fig:smallpop} shows 
that strong reciprocity, in the form of altruistic punishment,
can be selected for at 
the individual level in 
small populations ranging from $N=10$ to $N=10\,000$ agents when
selection is strong ($s=1$). The initial 
distribution of thresholds rapidly leads to a peaked function, with 
the range of acceptance thresholds for the agents covering about a 
10\% of the available ones. The position of the peak (understood as 
the mean acceptance threshold) fluctuates 
during the length of the simulation, never reaching a stationary value
for the durations we have explored. The width of the peak fluctuates 
as well, but in a much smaller scale than the position. 
At certain instants the distribution 
exhibits two peaks (see distribution at 7.5 million games). This is 
the mechanism by which the position of the peak moves around the 
possible acceptance thresholds. Importantly, the typical evolution 
we are describing does not depend on the initial condition. In particular,
a population consisting solely of self-interested agents, i.e., all 
initial thresholds are set to $t_i=1$, evolves in the same fashion.
The value $M$ of the capital at stake in every 
game is not important either, and increasing $M$ only leads 
to a higher resolution of the threshold distribution function.

\begin{figure}
\begin{center}
\includegraphics[height=14cm,angle=270]{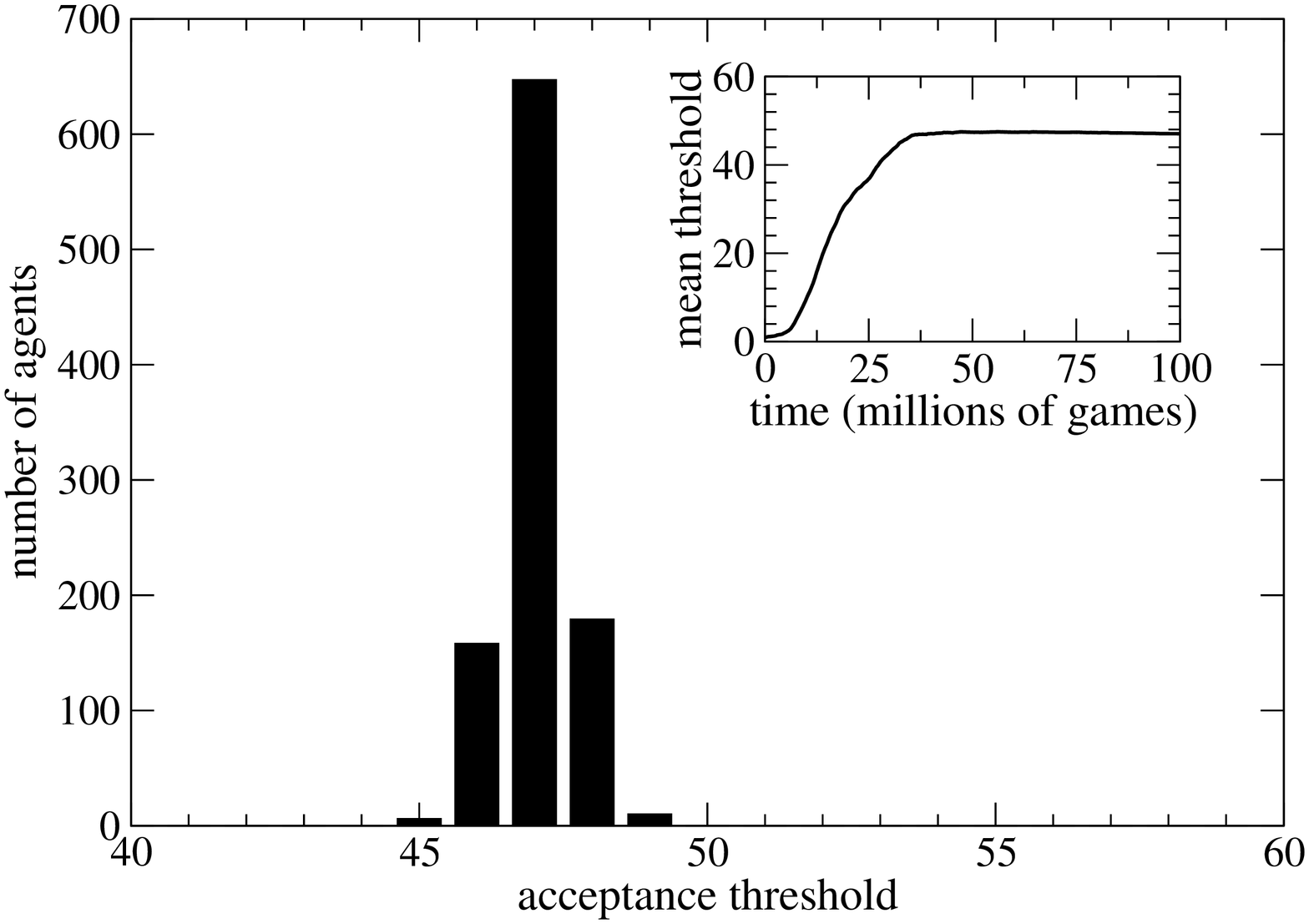}
\caption{\label{fig:slowsel}
}
\end{center}
\end{figure}

The success of reciprocators does not depend on the selection rate
(although the detailed dynamics does). 
Figure \ref{fig:slowsel} shows the result of a simulation with $1000$
agents in which the removal-duplication process takes place once every
$s=10\,000$ games.
To show further that the initial conditions are irrelevant, for this plot 
we have chosen an initial population of self-interested agents. As we may
see, the evolution is now much less noisy, and the distribution is narrower,
becoming highly peaked and immobile after a transient. The value of $s$ at 
which this regime appears increases with the population size. The final mean 
acceptance threshold at which simulations stabilize depends on the 
specific run, but it is very generally a value between 40 and 50. We thus
see that the selection rate may be responsible for the particulars of the
simulation outcome, but it is not a key factor for the emergence of 
strong reciprocity in our model.
%
%
{
We note, however, that taking very large values for $s$ or, strictly
speaking, considering the limit $s/N\to\infty$, does lead to different 
results. See next section for a detailed discussion.
}
%
%

\section{Discussion}

%
%
Among the results summarized above, the evolution of a population entirely 
formed by self-interested players into a diversified population with a 
large majority of altruists is the most relevant and surprising one. 
We will now argue that the underlying reason for this is the presence of
fluctuations (or noise) in our model. For the sake of definiteness, let us
consider the case $s=1$ (agent replacement takes place after every game) 
although the discussion applies to larger (but finite) values of $s$ as 
well. After one or more games, a mutation event will take place 
and a ``weak altruistic punisher'' (an agent with $t_i=2$) will appear
in the population, 
with a fitness inherited from its ancestor. For this new agent to be 
removed at the next iteration so that the population reverts to its 
uniform state, our model rules imply that this agent has to have 
the lowest fitness, that is the only one with that value of fitness,
{\em and also} that it does not play as a proposer in
the next game (if playing as a responder the agent will earn nothing
because of her threshold). In any other event this altruistic punisher
will survive at 
least one cycle, in which an additional one can appear by mutation.
%
%
Note also that in case a ``weak altruistic punisher'' is chosen to act
as a proposer, she earns a large amount of fitness, which would allow 
her to survive for many death-birth cycles, and during those she could 
even accumulate more fitness 
%
%
in case she is selected to play again as proposer. It is important to 
realize that this does not imply any constraint on the number of times
the emergent weak punisher is picked up as respondent: in that case,
and until a second punisher arises from mutation, acting as a respondent
the punisher will simply earn nothing, while the selfish agent playing
the role of proposer in that game would not earn the 99 fitness units
she would earn if she met another selfish agent. Therefore, the survival
of the first punisher does not depend on (and it may actually be favored
by) the number of times she acts as respondent, as one would expect in
a realistic situation.

The above discussion is in fact an example, something like a worst-case 
scenario for the $s=1$ case, and one can easily imagine other ways a 
newly created punisher may survive. Our intention is to illustrate
the crucial fact
%
%
%
%
%
%
that
%
%
fluctuations (i.e., the fact that the recently appeared
altruist is chosen to play or not, or that it is chosen to be removed if
there are more than one with the lowest fitness,
%
%
or other, selfish agents are not selected to play in one or several 
intervals)
%
%
allow for survival and
growth of the population of altruists. It is interesting to note that 
in the dynamics in which all players play against 
every other once, 
%
%
i.e., in the replicator dynamics (see next paragraph for more on this),
the average fitness earned by each type of agent can be computed 
analytically as a function of the frequency of the types in the 
population. From that result, it is easy to find out the threshold 
value required for one type to have a fitness advantage on the other. 
In particular, it can be shown that
%
%
if a 3\% of an initial $t_i=1$ population turns to $t_i=2$, 
the latter ones will outperform the originally self-interested agents.
Note also that, in our model, it can also be shown that the number of times
a particular agent is chosen to play is a random variable given by a 
Poisson distribution of mean $s/N$ (and of standard deviation 
$\sqrt{s/N}$, which for $s/N\gg 1$ becomes negligible with respect to the
mean).
Therefore, irrespective of their threshold, 
some agents will have played more than others and may have accumulated more
capital, subsequently being less exposed to removal.
All this scenario is what we refer to as `dynamics governed by fluctuations.' 

In the context of the above discussion, it is very illustrative to 
compare our results with 
%
%
previous studies of the Ultimatum Game by Page and
Nowak 
(Page and Nowak 2000, 2002).
The model introduced in those works has a dynamics completely 
different from ours: following standard evolutionary game theory,
every player plays every other one in both roles (proponent and 
respondent), and afterwards players reproduce with probability 
proportional to their payoff (which is fitness in the reproductive 
sense). Simulations and adaptive dynamics equations show then that the 
population ends up composed by players with fair (50\%) thresholds.
This is different from our observations, in which we hardly ever
reach an equilibrium (only for large $s$) and even then equilibria
set up at values different from the fair share. The reason for this
difference is that the Page-Nowak model dynamics describes the 
$s/N\to\infty$ limit of our model, in which between death-reproduction
events the time average gain all players obtain is
the mean payoff with high accuracy.
We thus see that our model is more general
%
%
{
because it has one free parameter, $s$, that allows selecting different
regimes whereas the Page-Nowak dynamics is only one limiting case.
Those different regimes are what we have described as fluctuation dominated 
(when $s/N$ is finite and not too large) and the regime analyzed by
Page and Nowak (when $s/N\to\infty$).
This amounts to saying that by varying $s$ we can
}
%
%
study regimes far from the standard evolutionary game theory 
limit. As a result, we find a variability of outcomes for the 
acceptance threshold consistent with the observations in real 
human societies %
(Fehr and Fischbacher, 2003; Gintis et al., 2003; Henrich et al., 2001).

\section{Two parameter model}

\label{sec:two}
To further confirm the differences between our approach and 
Page and Nowak's one, we have considered the same alternative as
they did, namely to assign agents 
a new strategical variable, $o_i$, defined as the amount
offered by player $i$ when acting as proponent, and subject to
the same mutation rules as the acceptance threshold, $t_i$. 
While Page and Nowak observed that in their setup, this 
modification of the model led to fully rational players
(i.e., in our model, $t_i=o_i=1$),
except for fluctuations due to mutations. Figure \ref{fig:two}
shows clearly that in our model the dynamics remains very 
complicated and equilibria are never reached within the 
duration of our simulations. Once again, this is due to the 
fact that the dynamics we propose does not remove the fluctuations
of the payoff obtained by the players as the limit $s/N\to\infty$ 
does.
%
%
{
It is clear that many other choices for the dynamics are possible,
aside from choosing different values for $s$. For instance, a certain 
percentage of the population could be replaced in reproduction 
events instead of just the least fit individual. Another possibility
would be the selection of individuals to be replaced with probability
given by their fitness. Notwithstanding, our main point here is that
our dynamics is far away from the replicator or adaptive ones, and 
the form we choose for the replacement is intend to make easier and 
faster to visualize the fluctuation effects. 
%
Our choice for the large
mutation rate points in the same direction as well, i.e., helps amplify 
the effect of fluctuations. In this respect, 
%
%
the question arises as to the influence of such a large mutation rate
in our results. To exclude any dependence of our main conclusion, namely
the appearance of altruistic punishers even in an initially selfish 
population, on the value of this quantity, we simulated the same model
for smaller mutation rates. 
Figure \ref{fig:mut} shows clearly that even for mutation rates as small 
as 1/3000 the population is taken over by the altruistic punishers, although
at a correspondingly larger time. 
Of course, the larger the mutation rate, the wider the histogram of the 
population, and therefore, for the smallest values the acceptance threshold
distribution is very sharply peaked around the mean value (see inset in
Figure 3). 
For even smaller rates,
(of the order of $10^{-4}$ or similar genetic 
mutation rates) 
the amount of time needed for 
altruistic individuals to establish becomes exceedingly large, and out
of the scope of our computing capabilities.
%
%
%
%
We believe that 
different rules for the dynamics would lead to qualitatively similar
results in so far as they do not approach Page and Nowak's (which 
we could call deterministic) limit. 
}
%
%

\section{Conclusions}

%
%
In this paper, we have shown that altruistic-like behavior, specifically,
altruistic punishment, may arise by means of exclusive individual selection
even in the absence of repeated interactions and reputation gains. Our 
conclusion is important in so far as it is generally believed that 
some kind of group selection is needed to understand the observed human
behavior. The reason for that is that game theoretical arguments 
apparently show that altruists are at disadvantage with respect to 
selfish individual. In this respect, another relevant conclusion
%
%
of the present work is that 
perspectives and approaches alternative to standard evolutionary 
game theory may be needed in order to understand paradoxical 
features such as the appearance of altruistic punishment. 
%
%
We begin by discussing this second conclusion and proceed to the 
first one afterwards. 
%
%

As we have seen, 
in our model the effects of finite time 
between generations 
%
%
{
(more precisely, the effect of keeping $s$ finite)
}
%
%
and of stochasticity play a non trivial 
role and sustain strong reciprocity (existence of players with
$t_i>1$) even if acceptance and offer obey independent rules.
%
%
{
Regarding this, it is important to notice that the way fluctuations enter
our model is directly through the evolutionary dynamics we propose. Other
important effects of noise have been reported in the literature 
(Gale et al., 1995; Binmore and Samuelson, 1999) in which fluctuations 
are included into a replicator dynamics for the Ultimatum game to 
account for imperfections in the learning process. In our case, there 
is no learning at all (agents have no memory) and therefore the source
of noise is the dynamics itself, i.e., the random differences between 
the number of games every player plays between selection events. 
Interestingly, randomness arising from finiteness of the population 
has also been shown to change the evolutionary 
}
%
%
stability of cooperation 
(Nowak et al., 2004).
In a related context, it has been 
recently reported that spatial structure, previously regarded as
beneficial for the evolution of cooperation on the basis of results
on the evolutionary Prisoner's Dilemma, may in fact inhibit it
(Hauert and Doebeli, 2004). Finally, let us also mention the 
recent results about the evolution of strong altruism in 
randomly formed groups when they exist for more than one generation
(Fletcher and Zwick, 2004).
All these unexpected and non trivial results, along with
our present 
report suggest that general approaches,
%
%
{
involving different, non-standard dynamics, 
}
%
%
beyond standard evolutionary
game theory, and particularly computer simulations of agent models,
may provide insights into the issue of how 
cooperation arises. Interestingly, it has been argued that empathy 
(or fairness),
i.e., the fact that agents offer what they themselves are prepared 
to accept, does not arise evolutionary on its own 
(Page and Nowak, 2002).
While those results are not questioned, they have been obtained in 
the framework of adaptive dynamics. We believe, along the same line
of reasoning we are presenting here, that the effect of fluctuations 
%
%
{
as described in the previous section
}
%
%
may be enough to originate and sustain fairness in finite populations,
which would in turn justify our model from the game theoretical 
viewpoint. 
%
%
In this regard, an interesting question arises when one considers the 
possibility of observing similar behavior in dilemma-type games 
(such as the prisonner's dilemma, see Axelrod and Hamilton, 1981). 
In that kind of games, the Nash equilibrium structure is much simpler
than in the Ultimatum game: Usually, they have only one equilibrium. 
It may then be that in those situations, departure from that equilibrium
by individual selection alone without additional ingredients is much 
more difficult. In other words, the existence of numerous Nash equilibria
in the Ultimatum game may facilitate the creative role of the fluctuations
in leading the population away from the self-interested type. It would 
be interesting to analyse the case of dilemma-type games in the light
of our findings here. 
%
%
Work along these lines is in progress.

Evolutionary explanations of strong reciprocity have been advanced in
terms of gene-culture coevolution 
(Bowles et al., 2003b; Bowles and Gintis, 2004;
Boyd et al., 2003; 
Gintis, 2003;
Hammerstein, 2003;
Henrich and Boyd, 2001).
The underlying rationale is that altruistic behavior leads to fitness
disadvantages at the individual level. But why must strong reciprocators
have lower fitness than other members of their group? While alternative
compensating factors (e.g., sexual selection) have been suggested 
(Bowles et al., 2003a),
our results show clearly that, in the context of the Ultimatum Game, 
altruistic punishment 
(Fehr and G\"achter, 2002)
may be established by individual 
selection alone. 
Our
simulations are consistent with the large degree of 
variability among individuals 
(Fehr and Fischbacher, 2003; Gintis et al., 2003)
and 
among societies 
(Henrich et al., 2001),
and reproduce the 
fact that typical offers are much larger than self-interested ones,
but lower than a fair share.
While in our
model agents have 
other-regarding
behavior (empathy), i.e., agents offer the minimum
they would accept if offered to them, this is not a requisite for 
the emergence of strong reciprocators 
as the two-threshold simulations show.
The population evolves by descent with modification and individual
selection, as the model does not implement cultural 
(other than parent-to-child transmission)
or group selection
of any kind.
To be sure, we do not mean that these mechanisms are 
irrelevant for the appearance and shaping of altruism: what we are 
showing is that strong reciprocity (and hence altruism)
may arise in their absence. Observations of strongly reciprocal 
behavior in capucin monkeys 
(Brosnan and de Waal, 2003),
where cultural transmission,
if any, is weak, strengthens this conclusion. Further support for our
thesis comes from reports of 
individual, pre-existent acceptance thresholds shown by 
neural activity measurements in 
(Sanfey et al., 2003).
In this respect, neural mechanisms gratifying cooperation as those 
demonstrated in 
(Rilling et al., 2002)
may have evolved to reinforce behaviors selected for at the individual 
level as we are suggesting.
%
%
{
Of course, those results do not preclude cultural influences in the 
brain control of altruistic behavior, which may play an important 
part in determining the experimentally observed thresholds. What is 
more relevant here is that individual thresholds do exist, with a 
large amount of individual variability, much like in our model, 
instead of a single culturally prescribed threshold. Evidence seems 
to favor one common threshold for acceptance and rejection but it is
not strong enough to exclude the other version of the model. 
}
%
%
The detrimental effects of unfair sanctions on altruism 
(Fehr and Rockenbach, 2003)
is yet another piece of evidence in favor of the existence of such 
individual acceptance (`fairness') thresholds.

In closing, let us emphasize that our
conclusion that altruism does not necessarily have negative
consequences for individuals draws such theories nearer to a biological
perspective.
Indeed, our results suggest that, 
despite its not being self-evident, altruistic strategies
may do better in terms of fitness than selfish ones, even 
without repeated interactions or reputation gain.
This conclusion, which would imply that strictly speaking there is no
truly altruistic behavior, may have
far-reaching implications in decision-making 
models and the design of public policies 
(Bowles et al., 2003a; Vogel, 2004).

\section*{Acknowledgments}

This work owes much to group discussions at GISC, for which we
thank its members, particularly Carlos Rasc\protect\'on for help with the literature.
A.S. is thankful to Maxi San Miguel for introducing him to the subject
and to Herbert Gintis for discussions.
We acknowledge financial support from Ministerio de Ciencia y Tecnolog\'\i a 
(Spain) through grants BFM2003-07749-C05-01 (AS) and BFM2003-0180 (JAC).

\section*{References}

\setlength{\parindent}{0cm}

Axelrod, R., Hamilton, W.\ D., 1981.
The evolution of cooperation.
Science {\bf 211}, 1390--1396. 

%
%
{
Binmore, K., Samuelson, L., 1999. 
Evolutionary drift and equilibrium selection. 
Rev.\ Econ.\ Stud.\ {\bf 66}, 363--393.
}
%
%
%

Bowles, S., Fehr, E., Gintis, H., 2003a.
Strong reciprocity may evolve
with or without group selection.
{\em Theoretical Primatology}, December issue. 

Bowles, S., Choi, J.-K., Hopfensitz, A., 2003b.
The co-evolution
of individual behaviors and social institutions.
J.\ Theor.\ Biol.\ {\bf 223}, 135--147.

Bowles S., Gintis, H., 2004.
The evolution of strong reciprocity:
cooperation in heterogeneous populations.
Theor.\ Popul.\ Biol.\ {\bf 65}, 17--28.

Boyd, R., Gintis, H., Bowles, S., Richerson, P.\ J., 2003. 
The
evolution of altruistic punishment.
Proc.\ Natl.\ Acad.\ Sci.\ USA {\bf 100}, 3531--3535.

Brosnan, S.\ F., de Waal, F.\ B.\ M., 2003. 
Monkeys reject unequal pay.
Nature {\bf 425}, 297--299.

Darwin, C., 1871. The Descent of Man, and Selection in 
Relation to Sex. Murray, London.

Fehr, E., Fischbacher, U., G\"achter, S., 2002.
Strong reciprocity, human
cooperation and the enforcement of social norms.
Hum.\ Nat.\ {\bf 13}, 1--25.

Fehr, E., Fischbacher, U., 2003.
The nature of human altruism.
Nature {\bf 425}, 785--791.

Fehr, E.\ G\"achter, S., 2002. 
Altruistic punishment
in humans.
Nature {\bf 415}, 137--140.

Fehr, E., Rockenbach, B., 2003. 
Detrimental effects of
sanctions on human altruism.
Nature {\bf 422}, 137--140.

Fletcher, J. A., Zwick, M., 2004. Strong altruism can evolve in 
randomly formed groups. J.\ Theor.\ Biol.\ {\bf 228}, 303--313.

%
%
{
Gale, J., Binmore, K., Samuelson, L., 1995. Learning to be 
imperfect: the ultimatum game. 
Games Econ.\ Behav.\ {\bf 8}, 56-90. 
}
%
%

%
%
Gintis, H. 2000. Game theory evolving. 
Princeton University Press, Princeton, NJ.
%
%

Gintis, H., 2000. 
Strong reciprocity and human sociality.
J.\ Theor.\ Biol.\ {\bf 206}, 169--179.

Gintis, H., 2003. The hitchhiker's guide to altruism: Gene-culture
co-evolution and the internalization of norms. J.\ Theor.\ Biol.\ {\bf 220},
407--418.

Gintis, H., Smith, E.\ A., Bowles, S., 2001.
Costly signalling and cooperation.
J.\ Theor. Biol.\ {\bf 213}, 103--119.

Gintis, H., Bowles, S., Boyd, R., Fehr, E., 2003.
Explaining altruistic behavior in humans.
Evol.\ Hum.\ Behav.\ {\bf 24}, 153--172.

Gould, S.\ J., 2002. The Structure of Evolutionary Theory.
Harvard University Press, Cambridge.

G\"uth, W., Schmittberger R., Schwarze, B., 1982. 
An experimental
analysis of ultimate bargaining.
J.\ Econ.\ Behav.\ Org.\ {\bf 3}, 367-388.

%
%
{
G\"uth, W., Schmidt, C., Sutter, M., 2003. 
Fairness in the mail and opportunism in the internet:  a newspaper
experiment on ultimatum bargaining. 
German Econ.\ Rev.\ {\bf 42}, 243-265. 
}
%
%

Hamilton, W.\ D., 1964.
The genetical evolution of social behavior (I and II).
J.\ Theor.\ Biol.\ {\bf 7}, 1--52.

Hammerstein, P., ed., 2003.
Genetic and Cultural Evolution of 
Cooperation. Dahlem Workshop Report 90. MIT Press, Cambridge, MA.

Hauert, C., Doebeli, M., 2004.
Spatial structure often inhibits the evolution of cooperation in 
the snowdrift game. 
Nature {\bf 428}, 643--646.

Henrich, J., Boyd, R., 2001. Why people punish defectors.
J.\ Theor.\ Biol.\ {\bf 208}, 79--89.

Henrich, J., Boyd, R., Bowles, S., Camerer, C.,
Fehr, E., Gintis, H., McElreath, R., 2001. 
In search of {\em Homo Economicus}:
Behavioral experiments in 15 small-scale societies.
Am.\ Econ.\ Rev.\ {\bf 91}, 73--78.

Leimar, O., Hammerstein, P., 2001. Evolution of cooperation through
indirect reciprocity. 
Proc.\ R.\ Soc.\ Lond.\ B {\bf 268}, 745--753.

Milinski, M., Semmann, D., Krambeck, H.\ J., 2002. 
Reputation
helps solve the `tragedy of the commons'.
Nature {\bf 415}, 424--426.

Nowak, M.\ A., Sigmund, K., 1998.
Evolution of indirect
reciprocity by image scoring.
Nature {\bf 393}, 573--577.

Nowak, M.\ A., Sasaki, A., Taylor, C.,
Fudenberg, D., 2004.
Emergence of cooperation and evolutionary stability in finite 
populations. 
Nature {\bf 428}, 646--650.

Page K.\ M., Nowak, M.\ A., 2000.
A generalized adaptive dynamics framework can describe the evolutionary
Ultimatum game. 
J.\ Theor.\ Biol.\ {\bf 209}, 173--179.

Page K.\ M., Nowak, M.\ A., 2002.
Empathy leads to fairness.
Bull.\ Math.\ Biol.\ {\bf 64}, 1101--1116.

Rilling, J.\ K., Gutman, D.\ A., Zeh, T.\ R., Pagnoni, G., 
Berns, G.\ S., Kilts, C.\ D., 2002.
A neural basis for cooperation.
Neuron, {\bf 35}, 395--405.

%
%
Samuelson, L., 1997. Evolutionary games and equilibrium 
selection. MIT Press, Cambridge, MA. 
%
%

Sanfey, A.\ G., Rilling, J.\ K., Aronson, J.\ A., 
Nystrom, L.\ E., Cohen, J.\ D., 2003. 
The neural basis of economic decision-making
in the ultimatum game.
Science {\bf 300}, 1755--1758.

Trivers, R.\ L., 1971.
The evolution of reciprocal altruism.
Q.\ Rev.\ Biol.\ {\bf 46}, 35--57.

Vogel, G., 2004.
The evolution of the golden rule.
Science {\bf 303}, 1128--1130.


\newpage

\section*{Figure captions}

\paragraph{Figure 1}
Non-self-interested behavior establishes spontaneosuly on small 
populations. Population size is $N=1000$, the capital to be shared
per game is $M=100$. Death and birth takes place after every 
game ($s=1$). Initial acceptance thresholds are distributed 
uniformly ($t_i=t_0$ conditions lead to the same output). 
Plotted are the distributions of acceptance threshold at the beginning
of the simulation and after 2, 7.5 and 10 million games.
Inset:
Mean acceptance threshold as a function of simulated time, 
is averaged over intervals of 10000 games to reduce noise
(in the raw data spikes appear that go above 50 or below 10).
The red
line in the inset is the average over all times of the mean,
located at 33.45. 

\paragraph{Figure 2}
Slow selection rates lead to stationary 
acceptance threshold distributions very narrowly peaked.
Population size is $N=1000$, the capital to be shared
per game is $M=100$ and selection is weak 
($s=10\,000$). Initial agents are all self-interested
($t_i=1$).
Plotted is the distribution of acceptance threshold at the end
of the simulation.
There are no agents with thresholds outside the plotted
range.  Inset:
Mean acceptance threshold as a function of simulated time. 
The asymptotically stable mean is very slowly approaching 47. 

\paragraph{Figure 3}
Introduction of an independent level for the amount of money 
offered by the agents does not change our conclusions. 
Population size is $N=1000$, the capital to be shared
per game is $M=100$ and selection is intermediate
($s=1000$). Initial agents are all fully rational
($t_i=o_i=1$).
Plotted are the distribution of acceptance threshold (ired)
and offered amount (green) after 50 million games (dashed)
and 100 million games (solid).
Upper inset: Mean acceptance threshold and offered amount as 
a function of simulated time. The offered amount is most of the
time larger than the acceptance threshold, and occasional 
crosses lead to a very slow dynamics until the situation 
is restored (see the plateaus around 62.5 million games, and
corresponding distributions in the lower inset).

\paragraph{Figure 4} 
Simulation results do not depend on the mutation rate. 
Population size is $N=10000$, the capital to be shared
per game is $M=100$. Death and birth takes place after every 
game ($s=1$).  Initial agents are all self-interested
($t_i=1$). Plotted is the mean value of the acceptance 
threshold vs time in number of games (which for $s=1$ 
equals the number of generations).
Mutation rates are as indicated in the plot.
Inset: Distribution of
the acceptance threshold at the end of each of the 
simulations. The fact that the peak of every distribution
is displaced to the left for larger mutation rates is 
an irrelevant coincidence: The peak positions fluctuate
in time (as indicated by the mean value in the main plot).

\newpage 

\newpage 

\newpage 

\begin{figure}
\begin{center}
\includegraphics[height=8cm,angle=0]{fig3.eps}\\[4mm]
\caption{\label{fig:two}
}
\end{center}
\end{figure}

\newpage 

\begin{figure}
\begin{center}
\includegraphics[height=8cm,angle=0]{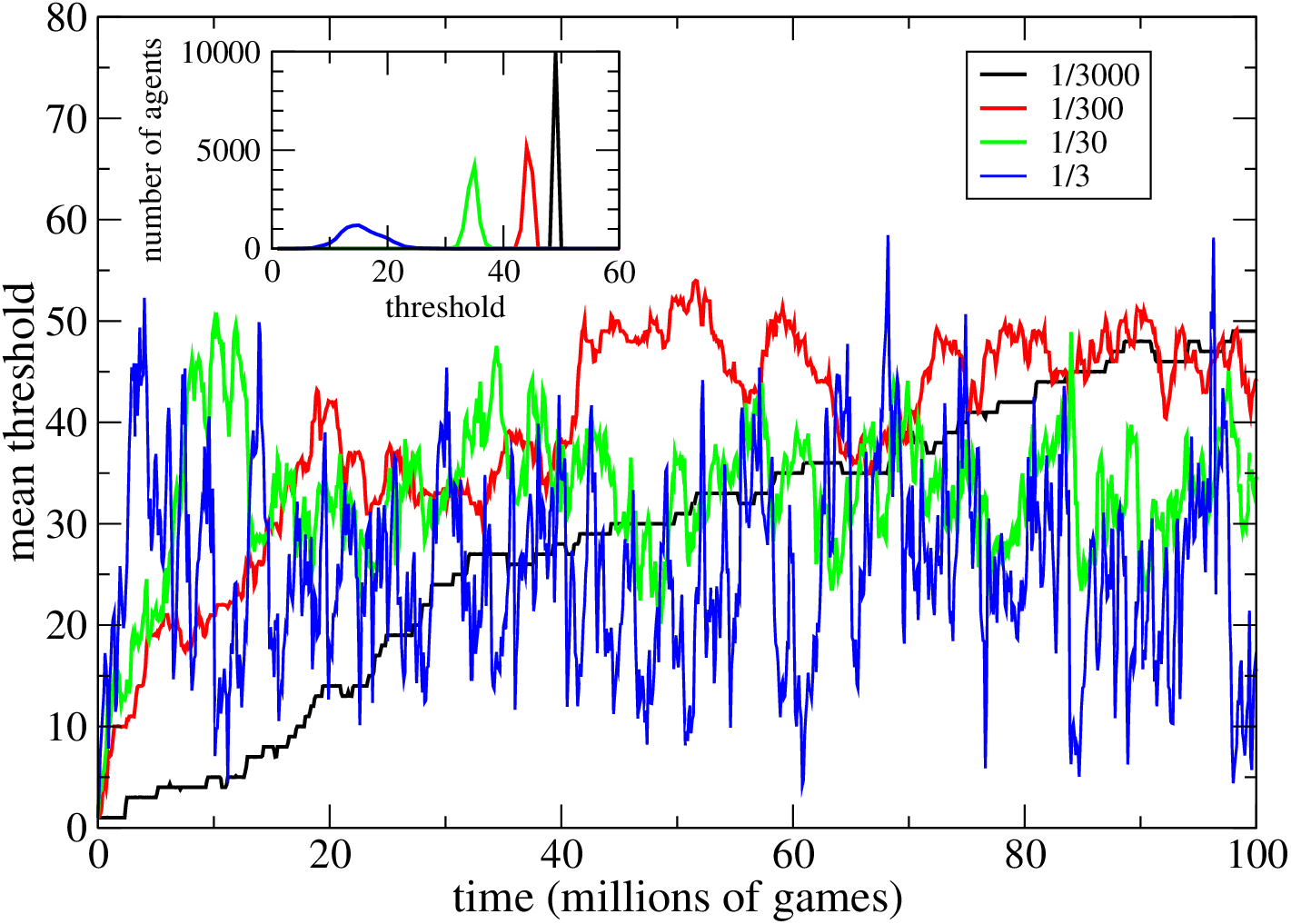}
\caption{\label{fig:mut}
}
\end{center}
\end{figure}

\end{document}